\documentclass[aps,prl,preprintnumbers,amsmath,amssymb,latexsym,nofootinbib,array,enumerate,letter,twocolumn,superscriptaddress]{revtex4}

\usepackage{graphicx}
\usepackage{epstopdf}
\usepackage{dcolumn}
\usepackage{bm}
\usepackage{hyperref}
\usepackage{color}
\usepackage{amsmath}

\begin{document}


\def\a{\alpha}
\def\b{\beta}
\def\c{\varepsilon}
\def\d{\delta}
\def\e{\epsilon}
\def\f{\phi}
\def\g{\gamma}
\def\h{\theta}
\def\k{\kappa}
\def\l{\lambda}
\def\m{\mu}
\def\n{\nu}
\def\p{\psi}
\def\q{\partial}
\def\r{\rho}
\def\s{\sigma}
\def\t{\tau}
\def\u{\upsilon}
\def\v{\varphi}
\def\w{\omega}
\def\x{\xi}
\def\y{\eta}
\def\z{\zeta}
\def\D{\Delta}
\def\G{\Gamma}
\def\H{\Theta}
\def\L{\Lambda}
\def\F{\Phi}
\def\P{\Psi}
\def\S{\Sigma}

\def\o{\over}
\def\beq{\begin{align}}
\def\eeq{\end{align}}
\newcommand{\gsim}{ \mathop{}_{\textstyle \sim}^{\textstyle >} }
\newcommand{\lsim}{ \mathop{}_{\textstyle \sim}^{\textstyle <} }
\newcommand{\vev}[1]{ \left\langle {#1} \right\rangle }
\newcommand{\bra}[1]{ \langle {#1} | }
\newcommand{\ket}[1]{ | {#1} \rangle }
\newcommand{\EV}{ {\rm eV} }
\newcommand{\KEV}{ {\rm keV} }
\newcommand{\MEV}{ {\rm MeV} }
\newcommand{\GEV}{ {\rm GeV} }
\newcommand{\TEV}{ {\rm TeV} }
\newcommand{\1}{\mbox{1}\hspace{-0.25em}\mbox{l}}
\newcommand{\headline}[1]{\noindent{\bf #1}}
\def\diag{\mathop{\rm diag}\nolimits}
\def\Spin{\mathop{\rm Spin}}
\def\SO{\mathop{\rm SO}}
\def\O{\mathop{\rm O}}
\def\SU{\mathop{\rm SU}}
\def\U{\mathop{\rm U}}
\def\Sp{\mathop{\rm Sp}}
\def\SL{\mathop{\rm SL}}
\def\tr{\mathop{\rm tr}}
\def\mpl{M_{\rm Pl}}

\def\IJMP{Int.~J.~Mod.~Phys. }
\def\MPL{Mod.~Phys.~Lett. }
\def\NP{Nucl.~Phys. }
\def\PL{Phys.~Lett. }
\def\PR{Phys.~Rev. }
\def\PRL{Phys.~Rev.~Lett. }
\def\PTP{Prog.~Theor.~Phys. }
\def\ZP{Z.~Phys. }

\def\dd{\mathrm{d}}
\def\ff{\mathrm{f}}
\def\BH{{\rm BH}}
\def\inf{{\rm inf}}
\def\ev{{\rm evap}}
\def\eq{{\rm eq}}
\def\SM{{\rm sm}}
\def\Mpl{M_{\rm Pl}}
\def\GeV{{\rm GeV}}
\newcommand{\Red}[1]{\textcolor{red}{#1}}
\newcommand{\TL}[1]{\textcolor{blue}{\bf TL: #1}}

\title{
QCD Axion Dark Matter with a Small Decay Constant 
}
\preprint{LCTP-17-11}

\author{Raymond T. Co}
\affiliation{Leinweber Center for Theoretical Physics, University of Michigan, Ann Arbor, Michigan 48109, USA}
\affiliation{Department of Physics, University of California, Berkeley, California 94720, USA}
\affiliation{Theoretical Physics Group, Lawrence Berkeley National Laboratory, Berkeley, California 94720, USA}
\author{Lawrence J. Hall}
\affiliation{Department of Physics, University of California, Berkeley, California 94720, USA}
\affiliation{Theoretical Physics Group, Lawrence Berkeley National Laboratory, Berkeley, California 94720, USA}
\author{Keisuke Harigaya}
\affiliation{Department of Physics, University of California, Berkeley, California 94720, USA}
\affiliation{Theoretical Physics Group, Lawrence Berkeley National Laboratory, Berkeley, California 94720, USA}

\begin{abstract}
The QCD axion is a good dark matter candidate.  The observed dark matter abundance can arise from misalignment or defect mechanisms, which generically require an axion decay constant $f_a \sim \mathcal{O}(10^{11})$ GeV (or higher). We introduce a new cosmological origin for axion dark matter, parametric resonance from oscillations of the Peccei-Quinn symmetry breaking field, that requires $f_a \sim (10^8 -10^{11})$ GeV.  The axions may be warm enough to give deviations from cold dark matter in Large Scale Structure.
\end{abstract}

\date{\today}

\maketitle

{\bf Introduction.}---%
The absence of CP violation from QCD is a long-standing problem in particle physics~\cite{tHooft:1976snw}
and is elegantly solved by the Peccei-Quinn (PQ) mechanism~\cite{Peccei:1977hh,Peccei:1977ur} involving a spontaneously broken anomalous symmetry. The scheme predicts the existence of a light boson~\cite{Weinberg:1977ma,Wilczek:1977pj}, the axion, which is constrained by data to be extremely light and hence is stable on cosmological time scales and a dark matter candidate.

What cosmological production mechanism yields axions that account for dark matter? Axions produced from the thermal bath have too low an abundance to be dark matter, and are too hot. 
Two production mechanisms have been considered; both are IR mechanisms with axions produced near the QCD phase transition.  While they depend on the PQ symmetry breaking scale, $f_a$, they are insensitive to details of the UV axion theory and dynamics of the PQ phase transition.  

\noindent i) Initially the axion is nearly massless, and takes a generic field value misaligned by angle $\theta_{\rm mis}$ from the vacuum value.
Around the QCD phase transition the axion obtains a mass, and the energy density in the resulting oscillations account for the dark matter abundance~\cite{Preskill:1982cy,Abbott:1982af,Dine:1982ah}
\begin{align}
\Omega_ah^2|_{\rm mis} \, \simeq \, & 0.01 \, \theta_{\rm mis}^2 \left(\frac{f_a}{10^{11}~{\rm GeV}}\right)^{1.19}.
\label{eq:mis}
\end{align}
\noindent ii) When PQ symmetry is broken after inflation, cosmic strings are produced~\cite{Kibble:1976sj}.  Domain walls form between these strings around the QCD phase transition and, if the domain wall number is unity, the string-domain wall network is unstable and decays into axions~\cite{Davis:1986xc},
yielding a dark matter density~\cite{Kawasaki:2014sqa,Klaer:2017ond}
\begin{align}
\Omega_ah^2|_{\rm string-DW} \,\simeq \, & 0.04 \mathchar`-0.3\left(\frac{f_a}{10^{11}~{\rm GeV}}\right)^{1.19}.
\label{eq:DW}
\end{align}

These mechanisms most naturally lead to axion dark matter for $f_a \sim (10^{11} - 10^{12})$ GeV, and the misalignment mechanism could also yield dark matter for larger $f_a$ as $\theta_{\rm mis}$ is reduced.  However, axions are not expected to be dark matter for smaller $f_a$~\footnote{The special cases for axion dark matter with lower $f_a$ are as follows. Refs.~\cite{Turner:1985si,Lyth:1991ub,Visinelli:2009zm} study the anharmonicity effects of the axion cosine potential when $\theta_{\rm mis}$ is tuned to approach $\pi$, requiring a small inflation scale to avoid quantum fluctuations. Ref.~\cite{Visinelli:2009kt} analyzes axion production in non-standard cosmological eras with kinetic energy domination even at temperatures below a GeV. Refs.~\cite{Hiramatsu:2010yn,Hiramatsu:2012sc,Kawasaki:2014sqa,Ringwald:2015dsf} consider topological defects for a domain wall number larger than unity with an explicit PQ breaking, where a fine tuning is required to solve the strong CP problem.}.

In this letter we introduce a mechanism for UV production of dark matter axions from the early evolution of the PQ symmetry breaking field $S$ in a relatively flat potential. We assume $S$ has a large initial value, $S_i \gg f_a$, e.g.~from a negative Hubble induced mass during inflation~\cite{Dine:1995uk}. After $S$ begins oscillating,
parametric resonance~\cite{Kofman:1994rk,Kofman:1997yn} creates a huge number of axions.  Assuming no subsequent entropy production
\begin{align}
\Omega_a h^2 \simeq 0.1 \left( \frac{S_i}{10^{16}~{\rm GeV}} \right)^2 \left( \frac{10~{\rm TeV}}{m_{S,i}} \right)^{ \scalebox{1.01}{$\frac{1}{2}$} } \frac{10^9~{\rm GeV}}{f_a},
\end{align}
where $m_{S,i}$ is the initial $S$ mass. Unlike (\ref{eq:mis}) and (\ref{eq:DW}), the abundance grows at low $f_a$ and sufficient dark matter can be obtained for $10^8 \, \mbox{GeV} < f_a < 10^{11} \, \mbox{GeV}$. Several experimental efforts are ongoing for axions with small $f_a$, including IAXO~\cite{Vogel:2013bta, Armengaud:2014gea} and TASTE~\cite{Anastassopoulos:2017kag} for solar axions, Orpheus \cite{Rybka:2014cya} and MADMAX~\cite{TheMADMAXWorkingGroup:2016hpc} for halo axions, and ARIADNE \cite{Arvanitaki:2014dfa, Geraci:2017bmq} for axion mediated CP-violating forces. See Refs.~\cite{Sikivie:2014lha, Arvanitaki:2017nhi, Baryakhtar:2018doz} for other proposals. Helioscopes will explore the lower end of this range, and it is worth noting that our mechanism allows values of $f_a$ even below $10^8$ GeV, which are also consistent with the supernova bound~\cite{Ellis:1987pk,Raffelt:1987yt,Turner:1987by,Mayle:1987as,Raffelt:2006cw} if there is a mild cancellation in the axion-nucleon coupling.

In our mechanism, axions are initially produced with momenta of order $m_{S,i}$. After subsequent red-shifting,
the axion velocity at a temperature of 1 eV is
\begin{equation}
v_a|_{\rm eV} \sim 10^{-4} \times \left( \frac{m_{S,i}}{10^6{\rm GeV}} \right)^{ \scalebox{1.01}{$\frac{1}{2}$} } \frac{f_a}{10^9~{\rm GeV}}.
\end{equation}
For a sufficiently large $m_{S,i}$, the axion is warm.

{\bf Axion production by oscillating PQ field.}---%
The axion mass is~\cite{Weinberg:1977ma}
\begin{equation}
m_a = 6~{\rm meV} \ \frac{10^9~{\rm GeV}}{f_a}.
\end{equation}
The axion number density $n_{a_0}$ that explains the observed dark matter abundance, $\rho_{\rm DM}/s \simeq 0.4~{\rm eV}$, is
\begin{equation}
\label{eq:Y_required}
Y_{a_0} \equiv \frac{n_{a_0}}{s}\simeq 70  \ \frac{f_a}{10^9~{\rm GeV}},
\end{equation}
where $s$ is the entropy density of the Universe. The required $Y_a$ is much larger than thermal for all $f_a \geq 10^8$ GeV, so axions must be produced non-thermally.

The radial direction of the PQ symmetry breaking field, which we call the saxion, $S$, (following terminology of supersymmetric theories) is taken to have a large initial field value.  As the Hubble scale $H$ becomes smaller than the saxion mass, the saxion begins coherent zero-mode oscillations.
The mass $m_S(S) \equiv \sqrt{V''(S)}$  is generally $S$ dependent and differs from the vacuum value.
For a wide range of potentials $V(S)$, the saxion yield is conserved with the initial value
\begin{align}
\label{eq:nSs}
Y_S \equiv \frac{n_S}{s} \sim Y_{S_i} \sim \frac{S_i^2}{ m_{S,i}^{1/2} \mpl^{3/2}},
\end{align}
in a radiation-dominated universe, where $\mpl \simeq 2.4 \times 10^{18}$ GeV.
A result for the matter-dominated case can be straightforwardly derived.
Suppose that this saxion yield is converted into axions,~e.g.~by decay.
For $ f_a \ll S_i < \mpl$, the dark matter axion yield of Eq.~(\ref{eq:Y_required}) results only if $m_{S,i} \ll S_i$, requiring $V(S)$ to be flat.

For a saxion to axion conversion rate $\Gamma_*$, at an oscillation amplitude $S_*$, the axion momentum $p_a(T) $ is
\begin{equation}
\frac{p_a(T)}{s^{1/3}} \simeq 0.5 \times \frac{m_{S*}}{\left(\Gamma_*\mpl\right)^{1/2}},
\end{equation}
assuming that axions are produced with momenta $\sim m_S(S_*) \equiv m_{S*}$.
Axions are colder for a larger $\Gamma_*$ since they receive more red-shifting after production.
Given the recent constraint on the warmness of dark matter (see e.g.~\cite{Irsic:2017ixq,Lopez-Honorez:2017csg}), we require that the axion velocity is smaller than $10^{-3}$ at a temperature of 1 eV,
\begin{equation}
\left. \frac{p_a(T)}{T} \right|_{T=1{\rm eV}} \, \lesssim 6 \times 10^{-6} \ \frac{10^9~{\rm GeV}}{f_a},
\label{eq:coldness}
\end{equation}
placing a lower bound on $\Gamma_*$.

Let us assume that saxion to axion conversion occurs via a perturbative decay rate of $\Gamma_* \sim m_{S_*}^3 / {S_+}^2$, where $S_+ = {\rm max}(S_*,f_a)$. The number density and momenta of the axions are given by
\begin{align}
\frac{n_a}{s} \sim \frac{S_*^2 S_+^3} {m_{S_*}^{7/2} \mpl^{3/2}},  \hspace{0.5in} \frac{p_a}{s^{1/3}}\sim \frac{S_+}{m_{S_*}^{1/2} \mpl^{1/2}}.
\end{align}
One can verify that the bound from coldness, (\ref{eq:coldness}), and the abundance requirement, (\ref{eq:Y_required}), are incompatible with each other for any $(S_*, m_{S_*})$, for
all $f_a$ larger than the experimental lower bound of $10^8$ GeV.

{\bf Parametric Resonance}---%
In fact, the production rate is typically much larger than the perturbative decay rate.
The saxion couples to the axion through the potential of the PQ breaking field, so that the axion mass oscillates.  Axion modes with non-zero momenta, which we call fluctuations, 
grow rapidly by parametric resonance~\cite{Kofman:1994rk,Kofman:1997yn} from initial seeds set by quantum fluctuations. The axion momentum is typically of order $m_S$. The energy of the fluctuations grows exponentially and at some point, typically soon after oscillations begin, becomes comparable to that of the saxion oscillation. At this stage, the back reaction on the saxion is non-negligible and parametric resonance ceases.   Due to efficient scattering between the oscillating saxion field and the axion fluctuations, the entire zero-mode energy of the saxion is converted into comparable saxion and axion fluctuations, $Y_S \sim Y_a \sim Y_{S_i}$, given in Eq.~(\ref{eq:nSs}). 

The evolution of the fluctuations after this stage is model-dependent, and a model-by-model lattice simulation is needed to rigorously follow the dynamics.
In this letter we consider the case where the co-moving momentum and number density of the fluctuations are approximately conserved,
so that the axion yield is comparable to that of the original saxion oscillation, given by Eq.~(\ref{eq:nSs}).

It is striking that this production mechanism, with $\Gamma \sim m_{S,i}$, occurs at a very early time. This
allows large red-shifting after production, giving axion momenta
\begin{equation}
\frac{p_a(T)}{s^{1/3}} \sim \left( \frac{m_{S,i}}{\mpl} \right)^{ \scalebox{1.01}{$\frac{1}{2}$} },
\end{equation}
that easily satisfy the coldness constraint (\ref{eq:coldness}) for sufficiently small saxion masses.

So far we assumed that entropy is conserved after axion production, but this depends on the fate of the saxions.  
To avoid overclosure, the energy of the saxions must be transferred to the thermal bath with some rate $\Gamma_s$ and hence thermalize at $T_{\rm th}$. However, thermalization after matter domination generates entropy, leading to an axion yield that is independent of $S_i$
\begin{align}
\label{eq:Ya_dom}
Y_a = \frac{n_a}{s} \sim \frac{T_{\rm th}}{m_{S,0}}, \hspace{0.5in} T_{\rm th} \sim \sqrt{\Gamma_s \mpl},
\end{align}
where $m_{S,0} \equiv m_S (f_a)$ is the vacuum saxion mass.

In the following, we apply the above mechanism to a few simple models of PQ symmetry breaking, giving results with and without matter domination by saxions.

Production of axions via parametric resonance is discussed in the literature. Refs.~\cite{Ema:2017krp, Ballesteros:2016euj, Ballesteros:2016xej} consider production of QCD axions by oscillations of the PQ symmetry breaking field, with axions identified as dark radiation. Ref.~\cite{Mazumdar:2015pta} investigates production of axion-like particle dark matter via derivative couplings with a scalar condensation. However, these papers do not consider production of QCD axion dark matter from parametric resonance.

{\bf Quartic potential.}---%
We consider a potential of the PQ symmetry breaking field $P$,
\begin{equation}
V = \lambda^2\left(|P|^2 - \frac{f_a^2}{2}\right)^2.
\end{equation}
The saxion mass is field-dependent, $m_S \simeq \sqrt{3}\lambda S $ for $S \gg f_a$ and $m_{S,0} = \sqrt{2} \lambda f_a$.
The saxion begins to oscillate when the Hubble scale is $\lambda S_i / \sqrt{3}\equiv H_{\rm osc}$.

After the saxion oscillation energy is transferred to axion and saxion fluctuations, the fluctuation momentum, initially of order $\lambda S$, slowly grows  via number-reducing scatterings as $k\propto R^{1/7}$~\cite{Micha:2002ey}  when the fluctuation amplitude is larger than $f_a$.  Here $R$ is the scale factor of the Universe. The evolution of the momentum for a smaller amplitude is not known, but we expect that the growth of the momentum becomes ineffective since the interaction term $\lambda^2 |P|^4$ is less effective.
The overall growth of the momentum is at the most  $\mathcal{O}(10)$ in the parameter space of interest, and we ignore it.
 
First we discuss the case of rapid thermalization of the saxion fluctuations so that they never dominate. The axion yield and momentum are given by
\begin{align}
\label{eq:Ya_4}
Y_a \simeq  \frac{0.04}{\lambda^{1/2}} \left(\frac{S_i}{ \mpl}\right)^{ \scalebox{1.01}{$\frac{3}{2}$} } \simeq 0.05 \left(\frac{S_i}{ \mpl}\right)^{ \scalebox{1.01}{$\frac{3}{2}$} } \left(\frac{f_a}{m_{S,0}}\right)^{ \scalebox{1.01}{$\frac{1}{2}$} } ,
\end{align}
\begin{align}
\label{eq:p_4}
\frac{p_a}{s^{1/3}} \simeq \lambda^{1/2}\left(\frac{S_i}{\mpl}\right)^{ \scalebox{1.01}{$\frac{1}{2}$} } \simeq \left(\frac{S_i}{\mpl}\right)^{ \scalebox{1.01}{$\frac{1}{2}$} } \left(\frac{m_{S,0}}{f_a}\right)^{ \scalebox{1.01}{$\frac{1}{2}$} },
\end{align}
giving constraints on the parameter space,
\begin{align}
\label{eq:quarticSi_lambda}
m_{S,0} \, \simeq & \; 400 \, {\rm MeV} \left( \frac{S_i}{2\times 10^{17} \, {\rm GeV}} \right)^3 \frac{10^9~{\rm GeV}}{f_a}, \nonumber\\
S_i \, \lesssim & \; 2 \times 10^{17} ~{\rm GeV}.
\end{align}

The axions should not be thermalized. For small $f_a$ thermalization seems inevitable, but in our mechanism it is suppressed by the large field value of $S$.
The axions scatter with the thermal bath with a rate 
\begin{align}
\Gamma_a \simeq 10^{-5} \frac{T^3}{ S_{{\rm eff}}^2 },  \hspace{0.5in} \frac{1}{S_{{\rm eff}}^2}\equiv  \vev{\frac{1}{2|P^2|}}.
\end{align}
During the zero-mode saxion oscillation $S_{\rm eff}$ may be much smaller than the amplitude of the oscillation as $P$ coherently gets small during the oscillation~\cite{Mukaida:2012qn}. However, $S_{\rm eff}$ is still large enough to prevent thermalization.
After the saxion oscillation energy is converted into fluctuations, $S_{{\rm eff}}$ is of order the amplitude of the fluctuation and decreases in proportion to the temperature, until it becomes of order $f_a$, after which $S_{{\rm eff}} = f_a$.  The ratio of the scattering rate to the Hubble scale is maximized when $S_{{\rm eff}}$ first reaches $f_a$.  Requiring this ratio to be smaller than unity gives
\begin{equation}
\label{eq:lambda_th}
m_{S,0} \lesssim 1 \, {\rm GeV} \left(\frac{S_i}{2 \times 10^{17}~{\rm GeV}}\right)^{ \scalebox{1.01}{$\frac{5}{2}$} } ,
\end{equation}
after removing $f_a$ using the constraint Eq.~(\ref{eq:quarticSi_lambda}) for the dark matter abundance.

\begin{figure}[t]
\begin{center}
\includegraphics[scale=0.44]{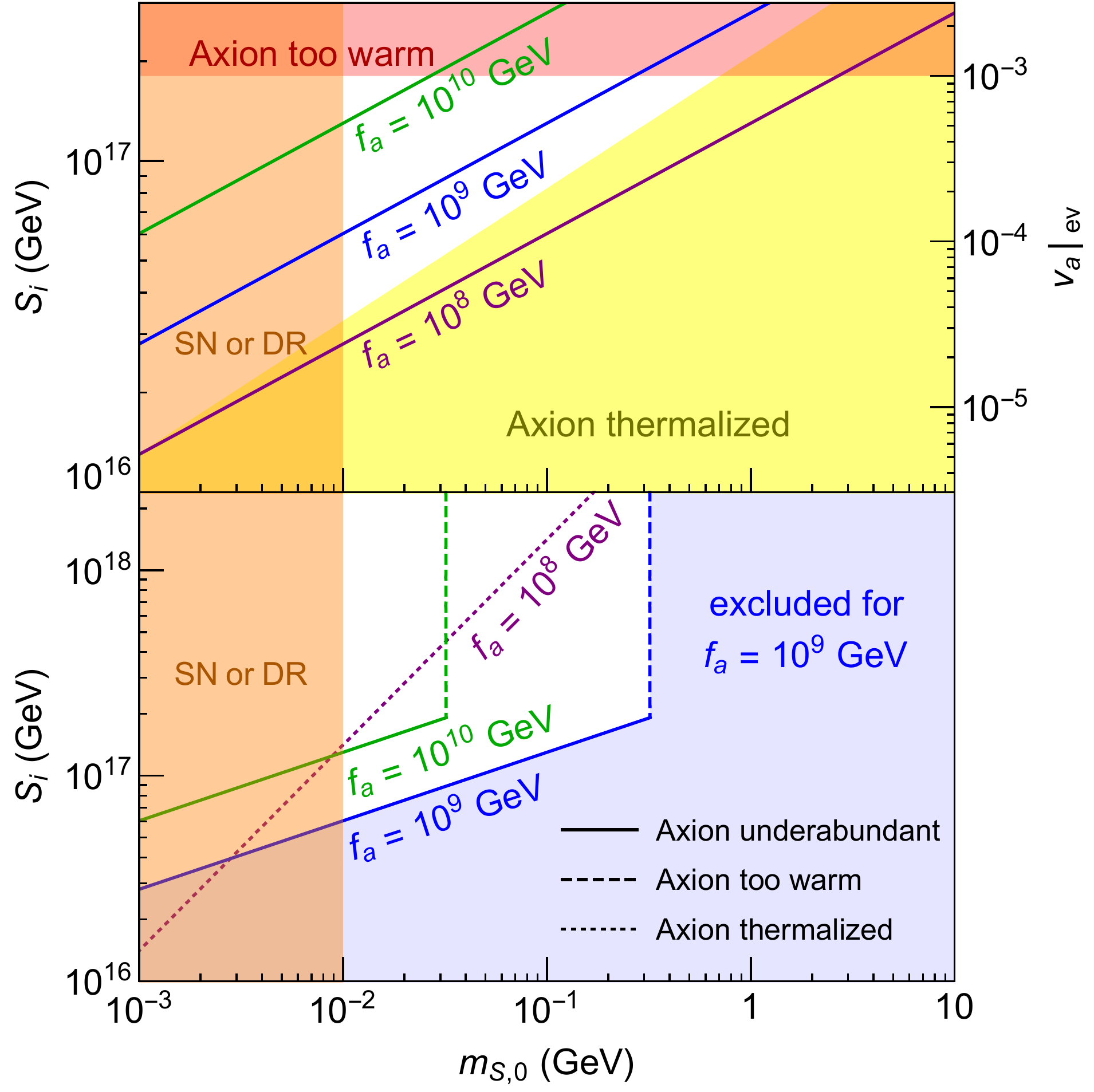}
\caption{
Constraints on the quartic theory in the parameter space of the saxion initial amplitude and vacuum mass, assuming saxion oscillations begin during a radiation-dominated era. The upper (lower) panel assumes an era of saxion domination does not (does) exist.  The solid lines of the two panels coincide; the observed dark matter abundance arises on (above) these lines for the upper (lower) panel.
}
\label{fig:quartic}
\end{center}
\end{figure}

The upper panel in Fig.~\ref{fig:quartic} summarizes the constraints on $m_{S,0}$ and $S_i$.
On each contour, dark matter is explained by axions produced from parametric resonance for the corresponding $f_a$. In the pink region, the axions are too warm to be dark matter; the y-axis on the right labels the axion velocity $v_a$ at $T=$ eV.
In the yellow region, the resonantly produced axions thermalize and are not dark matter.  The constraint leading to the orange region is explained later.

Once the zero-mode saxion oscillation is transferred into fluctuations of axions and saxions, their energy density evolves as radiation. After $S_{\rm eff}$ becomes of order $f_a$, the saxion behaves as matter, and may dominate the energy density of the universe.
The axion abundance is then given by Eq.~(\ref{eq:Ya_dom}), and the axion momentum is
\begin{align}
\frac{p_a}{s^{1/3}} \simeq \left( 36 \lambda^2  \frac{n_a}{s}\right)^{ \scalebox{1.01}{$\frac{1}{3}$} } = \left( 18 \frac{m_{S,0}^2}{f_a^2}  \frac{n_a}{s}\right)^{ \scalebox{1.01}{$\frac{1}{3}$} }.
\end{align}
With this dilution, the dark matter abundance requires the saxion mass to be smaller than in Eq.~(\ref{eq:quarticSi_lambda}).
The constraint from axion thermalization is derived in a similar manner to Eq.~(\ref{eq:lambda_th}), and is given by
\begin{equation}
m_{S,0} \lesssim 200 \, {\rm GeV} \left(\frac{f_a}{10^9~{\rm GeV}}\right)^3 \frac{S_i}{\mpl}.
\end{equation}
The constraints on parameters with a saxion domination era are shown in the lower panel of Fig.~\ref{fig:quartic}. The region to the right of each color line is excluded for the corresponding $f_a$. In regions between the orange boundary and the colored lines, the dark matter abundance is explained with $T_{\rm th}$ given by Eq.~(\ref{eq:Ya_dom}).

{\bf Quadratic potential.}---%
In supersymmetric QCD axion models, the saxion potential may be approximately quadratic. One example is a two-field model with superpotential and soft supersymmetry breaking terms
\begin{align}
W = X (P \bar{P} -f_a^2 ),~~V_{\rm soft} = m_P^2 |P|^2 + m_{\bar{P}}^2 |\bar{P}|^2.
\end{align}
Another example is a one-field model with a potential given by soft supersymmetry breaking and radiative corrections~\cite{Abe:2001cg,Nakamura:2008ey,DEramo:2015iqd},
\begin{equation}
\label{eq:potential_DT}
V = m^2 |P|^2 \left( {\rm ln} \frac{2 |P|^2}{f_a^2} -1 \right).
\end{equation}
For simplicity we approximate the dynamics of the saxion as the oscillation by an almost constant quadratic term with $m_{S,i} \simeq m_{S,0} \equiv m_S \sim m_P,m$.

In these models, the saxion starts oscillating when the Hubble scale is $H_{\rm osc} \simeq m_S/3$.
We parametrize the time of parametric resonance as $t\equiv N_p / m_S$.
After the zero-mode saxion oscillation energy is transferred to axion fluctuations,
the axion yield is of order that of the original saxion in Eq.~(\ref{eq:nSs}).
The axion momentum is
\begin{equation}
\frac{p_a}{s^{1/3}} \simeq 0.7 N_p^{1/2} \left(  \frac{m_S}{\mpl}\right)^{ \scalebox{1.01}{$\frac{1}{2}$} }.
\end{equation}
We have checked with a lattice simulation using LATTICEEASY~\cite{Felder:2000hq} that, in the one field model of Eq.~(\ref{eq:potential_DT}), the number density is conserved, the axion momentum does not grow due to number-reducing scatterings, and $N_p = \mathcal{O}(100)$. We discuss this issue in detail in a separate publication~\cite{CHH}. The results below also apply to the two-field theory if the comoving number density and momentum do not vary much.

We consider first the case that saxions thermalize before they dominate.  The abundance and coldness of axion dark matter then require
\begin{align}
\label{eq:quadraticSi_m}
m_S \, \simeq \, & \, 2~{\rm GeV} \left( \frac{S_i}{10^{15}~{\rm GeV}} \right)^4 \left( \frac{10^9~{\rm GeV}}{ f_a } \right)^2, \nonumber \\
S_i \lesssim \, & \; 3\times 10^{16}~{\rm GeV} \left( \frac{100}{N_p}  \right)^{1/4}.
\end{align}

The upper bound on $m_S$ from axion thermalization is
\begin{align}
m_S <
\left\{
\begin{array}{ll}
20~{\rm GeV} \left( \frac{f_a}{10^9~{\rm GeV}}  \right)^5 & : f_a < 3\times 10^9~{\rm GeV}, \\
2\times 10^4~{\rm GeV}   \frac{f_a}{10^{10}~{\rm GeV}}  & :  f_a > 3\times 10^9~{\rm GeV}.
\end{array}
\right.
\end{align}
The former is derived in a similar manner as for the quartic potential. When the saxion transfers its energy to the thermal bath before it dominates, $S_{\rm eff}$ suddenly drops to $f_a$, enhancing $\Gamma_a$. The latter bound is derived by conservatively assuming that saxions are destroyed at the temperature of saxion domination. If saxions are destroyed at higher temperatures, the constraint becomes stronger.
The bound is stronger than the one from the warmness of the axion only for $f_a \lesssim 10^9$ GeV.

\begin{figure}[tb]
\begin{center}
\includegraphics[scale=0.44]{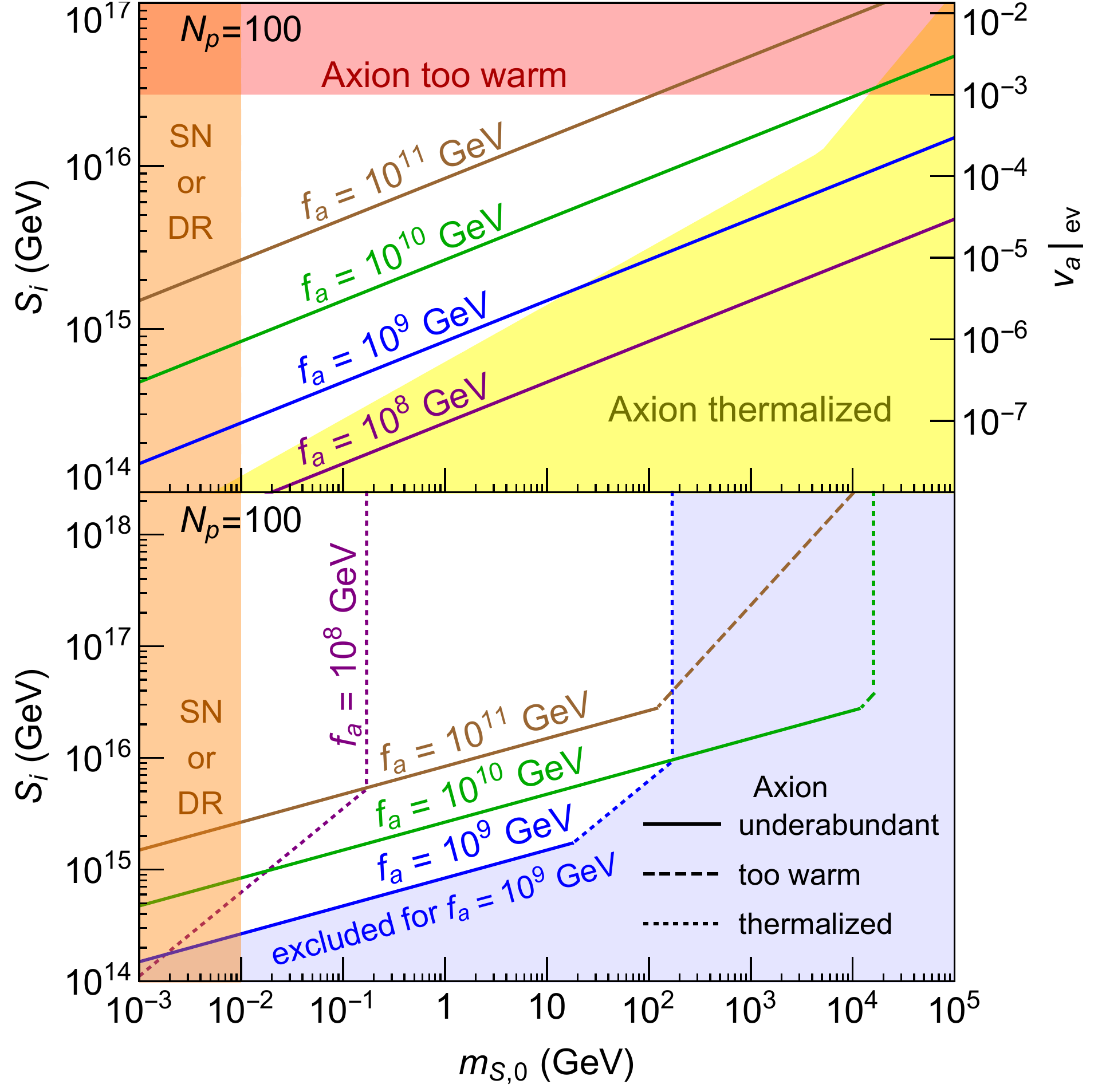}
\caption{
The same as Fig.~\ref{fig:quartic} but for the quadratic theory.
}
\label{fig:quadratic}
\end{center}
\end{figure}

In the case of an era of saxion domination, the axion abundance is given by Eq.~(\ref{eq:Ya_dom}), and the momentum is
\begin{equation}
\frac{p_a}{s^{1/3}} \simeq  \left( \frac{m_S^2 N_p^{3/2}}{S_i^2}  \frac{n_a}{s} \right)^{ \scalebox{1.01}{$\frac{1}{3}$} }.
\end{equation}
The saxion mass must be smaller than given in Eq.~(\ref{eq:quadraticSi_m}).
Axion thermalization places an upper bound on $m_S$,
\begin{align}
m_S < {\rm Min}
\left\{
\begin{array}{l}
3 \times 10^5~{\rm GeV} \left( \frac{f_a}{10^9~{\rm GeV}}  \right)^{ \frac{8}{3} } \left( \frac{S_i}{\mpl}  \right)^{ \frac{4}{3} }, \\
200 ~{\rm GeV}   \left( \frac{f_a}{10^9~{\rm GeV}} \right)^3.  \\ 
\end{array}
\right.
\end{align}
The former (latter) case is for thermalization during the radiation-dominated (saxion-dominated) era.
Formulae to derive the latter bound can be found in the appendix of Ref.~\cite{Co:2017pyf}.
Finally, when the saxion is destroyed $S_{\rm eff}$ suddenly drops to $f_a$, and the axion thermalization rate is enhanced, which gives an upper bound on $m_S$,
\begin{equation}
m_S < 2\times 10^4~{\rm GeV} \frac{f_a}{10^{10}~{\rm GeV}}.
\end{equation}

The constraints on the quadratic theory are shown in Fig.~\ref{fig:quadratic}.
In comparison with the quartic theory, a large vacuum mass of the saxion is allowed, since the mass is independent of the saxion field value.

For $S_i$ smaller than  $6\times 10^{15}$ GeV, saxions begin oscillating in the thermal potential generated by the free energy of quarks and gluons~\cite{Anisimov:2000wx}. (In principle, this can be avoided by giving a large mass to gluons.) This not only changes the saxion evolution, but may also lead to formation of clumpy objects such as Q-balls~\cite{Coleman:1985ki} and I-balls/oscillons~\cite{Bogolyubsky:1976nx,Gleiser:1993pt,Kasuya:2002zs}, which may drastically change axion production. We discuss this in a future work~\cite{CHH}.

{\bf The fate of the saxion.}---
Remnant saxions must be removed without causing cosmological problems.  If saxions, with yield (\ref{eq:Y_required}), do not decay or scatter with the thermal bath,
they dominate the energy density below temperature $T_{\rm dom}$,
\begin{equation}
\frac{T_{\rm dom}}{m_{S,0}} \simeq 100 \frac{f_a}{10^9~{\rm GeV}}.
\end{equation}
In this case, Eqs.~(\ref{eq:Y_required}) and (\ref{eq:Ya_dom}) imply $T_{\rm th} > m_{S,0}$.
Without saxion domination, the saxion must be thermalized at a temperature above $T_{\rm dom}$, larger than $m_{S,0}$. 
In both cases, $T_{\rm th} > m_{S,0}$ is required, and hence scattering with the thermal bath is important for destruction.
One may naively expect that scattering between axions and the thermal bath is also effective, thermalizing the axion.

This is, however, not always true.
Consider a coupling between $P$ and the Higgs field $H$,
$V = \kappa |P|^2 |H|^2$.
This induces a saxion-Higgs coupling,
$m_H^2 S |H|^2 / f_a$, but not an axion-Higgs coupling.
Here $m_H^2$ is a contribution to the Higgs mass squared from PQ symmetry breaking.
The saxion scattering rate with the thermal bath, before the electroweak phase transition, and for $T>m_S$, is~\cite{Mukaida:2012qn}
\begin{equation}
\Gamma_{s,H}\simeq \frac{1}{\pi} \frac{m_H^4 }{f_a^2 T} \left( \frac{S}{f_a} \right)^2.
\end{equation}
With this interaction the saxion is thermalized below a temperature $T_{\rm th}$,
\begin{align}
T_{\rm th} \simeq 1~{\rm TeV} \left( \frac{m_H}{200~{\rm GeV}} \right)^{ \scalebox{1.01}{$\frac{4}{3}$} }  \left(\frac{10^9~{\rm GeV}}{f_a} \right)^{ \scalebox{1.01}{$\frac{2}{3}$} },
\end{align}
which may be large enough.

Thermalized saxions interact with standard model fermions via mixing with the Higgs.
For sufficiently large $m_H$, saxions are still in thermal equilibrium while non-relativistic, and hence disappear without leaving any imprint by decaying into standard model fermions.  Saxion masses below $\mathcal{O}(10)$~MeV are however
excluded, for the $f_a$ of interest, from saxion cooling of supernovae~\cite{Ellis:1987pk,Raffelt:1987yt,Turner:1987by,Mayle:1987as,Raffelt:2006cw,Co:2017orl}.
For small $m_H$, on the other hand, saxions decouple from the thermal bath while relativistic, and decay dominantly to a pair of axions at temperature
\begin{align}
T_{\rm dec} \simeq 1~{\rm MeV} \left( \frac{m_{S,0}}{10~{\rm MeV}} \right)^{ \scalebox{1.01}{$\frac{3}{2}$} } \left( \frac{10^8~{\rm GeV}}{f_a} \right).
\end{align}
The resulting axions are observed as dark radiation of the Universe, with an energy density, normalized to one generation of neutrinos, given by
\begin{align}
& \Delta N_{\rm eff} = \frac{120 \zeta(3)}{7 \pi^4 }\frac{m_{S,0}}{T_{\rm dec}}  \frac{g_*(1 {\rm MeV})^{4/3}}{g_*(T_D) g_* (T_{\rm dec})^{1/3}}  \\
& \simeq   \, 0.3 \left(\frac{10~{\rm MeV}}{m_{S,0}}\right)^{ \scalebox{1.01}{$\frac{1}{2}$} }  \left( \frac{f_a}{10^8~{\rm GeV}} \right)   \frac{80}{g_* (T_{\rm D})} \left(\frac{10.75}{g_* (T_{\rm dec})}\right)^{ \scalebox{1.01}{$\frac{1}{3}$} } \nonumber,
\end{align}
where $g_*(T)$ is the number of degrees of freedom of the thermal bath at temperature $T$, and $T_{D}$ is the saxion decoupling temperature.
The experimental upper bound is $\Delta N_{\rm eff} < 0.6$~\cite{Ade:2015xua}.
For $m_{S,0} < 10$ MeV, there is too much dark radiation (small $m_H$) or too rapid supernova cooling (large $m_H$), as shown by the orange region in Figs.~\ref{fig:quartic} and \ref{fig:quadratic}.

 {\bf Discussion.}---%
In a wide class of theories, a large initial value of the PQ breaking field leads to cosmological production of axions via parametric resonance. 
The resulting dark matter abundance scales as $1/f_a$ and dominates at low $f_a$.  Figures 1 and 2 show that, in theories with quartic and quadratic potentials, there are large regions of parameter space allowing axion dark matter with $f_a \sim (10^8 - 10^{11})$ GeV.  The parts of these regions close to the ``too warm" boundaries have Large Scale Structure that deviates from that of cold dark matter.  The required flatness of the potential leads to a mass hierarchy, $m_{S,0} \ll f_a$, that can be understood with supersymmetry.

Parametric resonance randomizes the axion direction so that axions are also produced from the mis-alignment mechanism with $\theta_{\rm mis}= \mathcal{O}(1)$, overproducing axions for $f_a \gtrsim 10^{12}$ GeV.  
In the quartic theory, PQ symmetry is restored, leading to domain wall formation~\cite{Tkachev:1995md,Kasuya:1996ns}. This might also occur in the one-field quadratic model. The domain wall number should be unity so that domain walls are unstable~\cite{Sikivie:1982qv}. The unstable string domain wall network overproduces axions if $f_a \gtrsim 10^{12}$ GeV. A numerical simulation~\cite{Kawasaki:2017kkr} suggests that domain walls are not produced in the two-field quadratic model.
 
If the reheating temperature of the Universe is low enough, oscillations begin during an inflaton-dominated era;
this case will be discussed in a future publication, along with a detailed analysis on thermal potentials~\cite{CHH}.
 
 {\bf Acknowledgment.}---%
The authors thank Kyohei Mukaida and Yue Zhao for discussions.
This work was supported in part by the Director, Office of Science, Office of High Energy and Nuclear Physics, of the US Department of Energy under Contract DE-AC02-05CH11231 and by the National Science Foundation under grants PHY-1316783 and PHY-1521446.


\begin{thebibliography}{99} 

\bibitem{tHooft:1976snw} 
  G.~'t Hooft,
  Phys.\ Rev.\ D {\bf 14}, 3432 (1976)
  Erratum: [Phys.\ Rev.\ D {\bf 18}, 2199 (1978)].

\bibitem{Peccei:1977hh} 
  R.~D.~Peccei and H.~R.~Quinn,
  Phys.\ Rev.\ Lett.\  {\bf 38}, 1440 (1977).

\bibitem{Peccei:1977ur} 
  R.~D.~Peccei and H.~R.~Quinn,
  Phys.\ Rev.\ D {\bf 16}, 1791 (1977).

\bibitem{Weinberg:1977ma} 
  S.~Weinberg,
  Phys.\ Rev.\ Lett.\  {\bf 40}, 223 (1978).

\bibitem{Wilczek:1977pj} 
  F.~Wilczek,
  Phys.\ Rev.\ Lett.\  {\bf 40}, 279 (1978).

\bibitem{Preskill:1982cy} 
  J.~Preskill, M.~B.~Wise and F.~Wilczek,
  Phys.\ Lett.\  {\bf 120B}, 127 (1983).

\bibitem{Abbott:1982af} 
  L.~F.~Abbott and P.~Sikivie,
  Phys.\ Lett.\  {\bf 120B}, 133 (1983).
   
\bibitem{Dine:1982ah} 
  M.~Dine and W.~Fischler,
  Phys.\ Lett.\  {\bf 120B}, 137 (1983).

\bibitem{Kibble:1976sj} 
  T.~W.~B.~Kibble,
  J.\ Phys.\ A {\bf 9}, 1387 (1976).

\bibitem{Davis:1986xc} 
  R.~L.~Davis,
  Phys.\ Lett.\ B {\bf 180}, 225 (1986).

\bibitem{Kawasaki:2014sqa} 
  M.~Kawasaki, K.~Saikawa and T.~Sekiguchi,
  Phys.\ Rev.\ D {\bf 91}, no. 6, 065014 (2015)
  [arXiv:1412.0789 [hep-ph]].

\bibitem{Klaer:2017ond} 
  V.~B.~Klaer and G.~D.~Moore,
  arXiv:1708.07521 [hep-ph].

\bibitem{Turner:1985si} 
  M.~S.~Turner,
  Phys.\ Rev.\ D {\bf 33}, 889 (1986).

\bibitem{Lyth:1991ub} 
  D.~H.~Lyth,
  Phys.\ Rev.\ D {\bf 45}, 3394 (1992).

\bibitem{Visinelli:2009zm} 
  L.~Visinelli and P.~Gondolo,
  Phys.\ Rev.\ D {\bf 80}, 035024 (2009)
  [arXiv:0903.4377 [astro-ph.CO]].
  
\bibitem{Visinelli:2009kt} 
  L.~Visinelli and P.~Gondolo,
  Phys.\ Rev.\ D {\bf 81}, 063508 (2010)
  [arXiv:0912.0015 [astro-ph.CO]].

\bibitem{Hiramatsu:2010yn} 
  T.~Hiramatsu, M.~Kawasaki and K.~Saikawa,
  JCAP {\bf 1108}, 030 (2011)
  [arXiv:1012.4558 [astro-ph.CO]].

\bibitem{Hiramatsu:2012sc} 
  T.~Hiramatsu, M.~Kawasaki, K.~Saikawa and T.~Sekiguchi,
  JCAP {\bf 1301}, 001 (2013)
  [arXiv:1207.3166 [hep-ph]].

\bibitem{Ringwald:2015dsf} 
  A.~Ringwald and K.~Saikawa,
  Phys.\ Rev.\ D {\bf 93}, no. 8, 085031 (2016)
  Addendum: [Phys.\ Rev.\ D {\bf 94}, no. 4, 049908 (2016)]
  [arXiv:1512.06436 [hep-ph]].

\bibitem{Dine:1995uk} 
  M.~Dine, L.~Randall and S.~D.~Thomas,
  Phys.\ Rev.\ Lett.\  {\bf 75}, 398 (1995)
  [hep-ph/9503303].

\bibitem{Kofman:1994rk} 
  L.~Kofman, A.~D.~Linde and A.~A.~Starobinsky,
  Phys.\ Rev.\ Lett.\  {\bf 73}, 3195 (1994)
  [hep-th/9405187].

\bibitem{Kofman:1997yn} 
  L.~Kofman, A.~D.~Linde and A.~A.~Starobinsky,
  Phys.\ Rev.\ D {\bf 56}, 3258 (1997)
  [hep-ph/9704452].

\bibitem{Vogel:2013bta} 
  J.~K.~Vogel {\it et al.},
  arXiv:1302.3273 [physics.ins-det].

\bibitem{Armengaud:2014gea} 
  E.~Armengaud {\it et al.},
  JINST {\bf 9}, T05002 (2014)
  [arXiv:1401.3233 [physics.ins-det]].

\bibitem{Anastassopoulos:2017kag} 
  V.~Anastassopoulos {\it et al.} [TASTE Collaboration],
  arXiv:1706.09378 [hep-ph].

\bibitem{Rybka:2014cya} 
  G.~Rybka, A.~Wagner, A.~Brill, K.~Ramos, R.~Percival and K.~Patel,
  Phys.\ Rev.\ D {\bf 91}, no. 1, 011701 (2015)
  [arXiv:1403.3121 [physics.ins-det]].

\bibitem{TheMADMAXWorkingGroup:2016hpc} 
  A.~Caldwell {\it et al.} [MADMAX Working Group],
  Phys.\ Rev.\ Lett.\  {\bf 118}, no. 9, 091801 (2017)
  [arXiv:1611.05865 [physics.ins-det]].
  
\bibitem{Arvanitaki:2014dfa} 
  A.~Arvanitaki and A.~A.~Geraci,
  Phys.\ Rev.\ Lett.\  {\bf 113}, no. 16, 161801 (2014)
  [arXiv:1403.1290 [hep-ph]].
  
\bibitem{Geraci:2017bmq} 
  A.~A.~Geraci {\it et al.},
  arXiv:1710.05413 [astro-ph.IM].

\bibitem{Sikivie:2014lha} 
  P.~Sikivie,
  Phys.\ Rev.\ Lett.\  {\bf 113}, no. 20, 201301 (2014)
  [arXiv:1409.2806 [hep-ph]].
  
\bibitem{Arvanitaki:2017nhi} 
  A.~Arvanitaki, S.~Dimopoulos and K.~Van Tilburg,
  arXiv:1709.05354 [hep-ph].
  
\bibitem{Baryakhtar:2018doz} 
  M.~Baryakhtar, J.~Huang and R.~Lasenby,
  arXiv:1803.11455 [hep-ph].
  
\bibitem{Ellis:1987pk} 
  J.~R.~Ellis and K.~A.~Olive,
    Phys.\ Lett.\ B {\bf 193}, 525 (1987)

\bibitem{Raffelt:1987yt} 
  G.~Raffelt and D.~Seckel,
  Phys.\ Rev.\ Lett.\  {\bf 60}, 1793 (1988).

\bibitem{Turner:1987by} 
  M.~S.~Turner,
  Phys.\ Rev.\ Lett.\  {\bf 60}, 1797 (1988).

\bibitem{Mayle:1987as} 
  R.~Mayle, J.~R.~Wilson, J.~R.~Ellis, K.~A.~Olive, D.~N.~Schramm and G.~Steigman,
  Phys.\ Lett.\ B {\bf 203}, 188 (1988).

\bibitem{Raffelt:2006cw} 
  G.~G.~Raffelt,
  Lect.\ Notes Phys.\  {\bf 741}, 51 (2008)
  [hep-ph/0611350].

\bibitem{Irsic:2017ixq} 
  V.~Ir$\check{\rm s}$i$\check{\rm c}$ {\it et al.},
  Phys.\ Rev.\ D {\bf 96}, no. 2, 023522 (2017)
  [arXiv:1702.01764 [astro-ph.CO]].

\bibitem{Lopez-Honorez:2017csg} 
  L.~Lopez-Honorez, O.~Mena, S.~Palomares-Ruiz and P.~V.~Domingo,
  arXiv:1703.02302 [astro-ph.CO].


%
%
%

\bibitem{Ema:2017krp} 
  Y.~Ema and K.~Nakayama,
  arXiv:1710.02461 [hep-ph].

\bibitem{Ballesteros:2016euj} 
  G.~Ballesteros, J.~Redondo, A.~Ringwald and C.~Tamarit,
  Phys.\ Rev.\ Lett.\  {\bf 118}, no. 7, 071802 (2017)
  [arXiv:1608.05414 [hep-ph]].
  
\bibitem{Ballesteros:2016xej} 
  G.~Ballesteros, J.~Redondo, A.~Ringwald and C.~Tamarit,
  JCAP {\bf 1708}, no. 08, 001 (2017)
  [arXiv:1610.01639 [hep-ph]].

\bibitem{Mazumdar:2015pta} 
  A.~Mazumdar and S.~Qutub,
  Phys.\ Rev.\ D {\bf 93}, no. 4, 043502 (2016)
  [arXiv:1508.04136 [hep-ph]].


\bibitem{Micha:2002ey} 
  R.~Micha and I.~I.~Tkachev,
  Phys.\ Rev.\ Lett.\  {\bf 90}, 121301 (2003)
  [hep-ph/0210202].


\bibitem{Mukaida:2012qn} 
  K.~Mukaida and K.~Nakayama,
  JCAP {\bf 1301}, 017 (2013)
  [arXiv:1208.3399 [hep-ph]].



%





\bibitem{Abe:2001cg} 
  N.~Abe, T.~Moroi and M.~Yamaguchi,
  JHEP {\bf 0201}, 010 (2002)
  [hep-ph/0111155].

\bibitem{Nakamura:2008ey} 
  S.~Nakamura, K.~i.~Okumura and M.~Yamaguchi,
  Phys.\ Rev.\ D {\bf 77}, 115027 (2008)
  [arXiv:0803.3725 [hep-ph]].

\bibitem{DEramo:2015iqd} 
  F.~D'Eramo, L.~J.~Hall and D.~Pappadopulo,
  JHEP {\bf 1506}, 117 (2015)
  [arXiv:1502.06963 [hep-ph]].

\bibitem{Felder:2000hq} 
  G.~N.~Felder and I.~Tkachev,
  Comput.\ Phys.\ Commun.\  {\bf 178}, 929 (2008)
  [hep-ph/0011159].

\bibitem{CHH}
  R.~T.~Co, L.~J.~Hall and K.~Harigaya, in preparation.


\bibitem{Co:2017pyf} 
  R.~T.~Co and K.~Harigaya,
  JHEP {\bf 1710}, 207 (2017)
  [arXiv:1707.08965 [hep-ph]].


\bibitem{Anisimov:2000wx} 
  A.~Anisimov and M.~Dine,
  Nucl.\ Phys.\ B {\bf 619}, 729 (2001)
  [hep-ph/0008058].
  
\bibitem{Coleman:1985ki} 
  S.~R.~Coleman,
  Nucl.\ Phys.\ B {\bf 262}, 263 (1985)
  Erratum: [Nucl.\ Phys.\ B {\bf 269}, 744 (1986)].

\bibitem{Bogolyubsky:1976nx} 
  I.~L.~Bogolyubsky and V.~G.~Makhankov,
  JETP Lett.\  {\bf 24}, 12 (1976).

\bibitem{Gleiser:1993pt} 
  M.~Gleiser,
  Phys.\ Rev.\ D {\bf 49}, 2978 (1994)
  [hep-ph/9308279].

\bibitem{Kasuya:2002zs} 
  S.~Kasuya, M.~Kawasaki and F.~Takahashi,
  Phys.\ Lett.\ B {\bf 559}, 99 (2003)
  [hep-ph/0209358].
  

\bibitem{Co:2017orl} 
  R.~T.~Co, F.~D'Eramo, L.~J.~Hall and K.~Harigaya,
  JHEP {\bf 1707}, 125 (2017)
  [arXiv:1703.09796 [hep-ph]].

\bibitem{Ade:2015xua} 
  P.~A.~R.~Ade {\it et al.} [Planck Collaboration],
  Astron.\ Astrophys.\  {\bf 594}, A13 (2016)
  [arXiv:1502.01589 [astro-ph.CO]].

\bibitem{Tkachev:1995md} 
  I.~I.~Tkachev,
  Phys.\ Lett.\ B {\bf 376}, 35 (1996)
  [hep-th/9510146].

\bibitem{Kasuya:1996ns} 
  S.~Kasuya, M.~Kawasaki and T.~Yanagida,
  Phys.\ Lett.\ B {\bf 409}, 94 (1997)
  [hep-ph/9608405].

\bibitem{Sikivie:1982qv} 
  P.~Sikivie,
  Phys.\ Rev.\ Lett.\  {\bf 48}, 1156 (1982).

\bibitem{Kawasaki:2017kkr} 
  M.~Kawasaki and E.~Sonomoto,
  arXiv:1710.07269 [hep-ph].


\end{thebibliography}
\end{document}